\documentclass[submission, Proceedings]{SciPost}

\hypersetup{
    colorlinks,
    linkcolor={red!50!black},
    citecolor={blue!50!black},
    urlcolor={blue!80!black}
}

\usepackage[bitstream-charter]{mathdesign}
\usepackage{caption}
\usepackage{subcaption}
\usepackage{tikz}
\usepackage{siunitx}
\usepackage{booktabs}
\usepackage{multirow}
\usepackage[capitalize]{cleveref}
\newcommand{\ptl}{p_\perp^\ell}
\newcommand{\avgptl}[1]{\langle p_\perp^{\ell,#1} \rangle}
\newcommand{\rmd}{\mathrm{d}}

\DeclareSymbolFont{usualmathcal}{OMS}{cmsy}{m}{n}
\DeclareSymbolFontAlphabet{\mathcal}{usualmathcal}

\begin{document}

\begin{center}{\Large \textbf{
Mixed QCD-electroweak corrections to $Z$ and $W$ boson production
and their impact on the $W$ mass measurements at the LHC \\
}}\end{center}

\begin{center}
A. Behring\textsuperscript{1$\star$}
\end{center}

\begin{center}
{\bf 1} Institute for Theoretical Particle Physics, KIT, Karlsruhe, Germany
\\
* \href{mailto:arnd.behring@cern.ch}{arnd.behring@cern.ch}
\end{center}

\begin{center}
\today
\end{center}


\definecolor{palegray}{gray}{0.95}
\begin{center}
\colorbox{palegray}{
  \begin{tabular}{rr}
  \begin{minipage}{0.1\textwidth}
    \includegraphics[width=35mm]{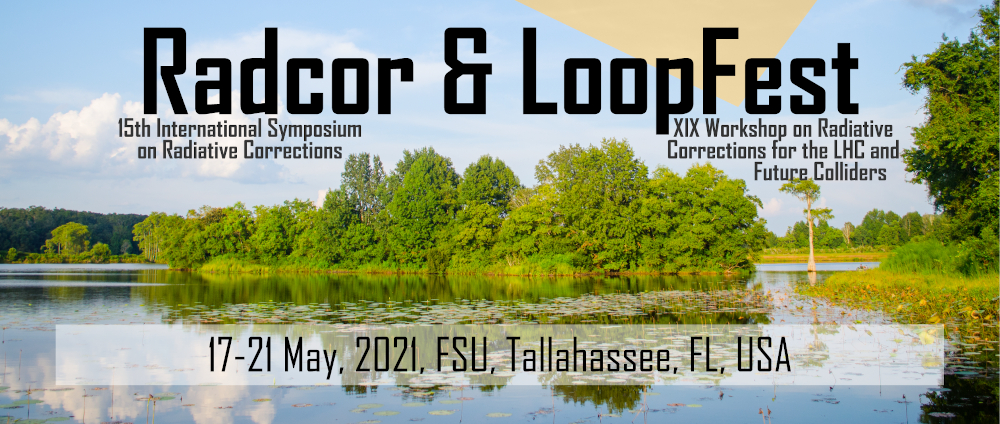}
  \end{minipage}
  &
  \begin{minipage}{0.85\textwidth}
    \begin{center}
    {\it 15th International Symposium on Radiative Corrections: \\Applications of Quantum Field Theory to Phenomenology,}\\
    {\it FSU, Tallahasse, FL, USA, 17-21 May 2021} \\
    \doi{10.21468/SciPostPhysProc}\\
    \end{center}
  \end{minipage}
\end{tabular}
}
\end{center}

\section*{Abstract}
{\bf
We report on recently computed mixed QCD-electroweak corrections to
on-shell $W$ and $Z$ boson production. We use these differential
predictions to estimate their impact on the $W$ boson mass
determination at the LHC.
}

\section{Introduction}
\label{sec:intro}
The mass of the $W$ boson, $m_W$, has been measured with very high precision at
both lepton and hadron colliders. The most recent measurement by the ATLAS
collaboration \cite{ATLAS:2017rzl} quotes an uncertainty of $\SI{19}{\MeV}$,
which can be compared to $\SI{8}{\MeV}$ uncertainty from global electroweak fits
\cite{Baak:2014ora,deBlas:2016ojx}. A very precise knowledge of this and other
Standard Model parameters allows to cross-check the internal consistency of the
Standard Model and to search for hints of new physics. CMS and ATLAS aim to
further reduce the uncertainty to about $\mathcal{O}(\SI{10}{\MeV})$ which would
rival the precision from global electroweak fits. This corresponds to an
astounding precision of about $0.01\%$.

In order to measure the $W$ boson mass at a hadron collider, one can use
on-shell production of a single vector boson and its subsequent decay into
leptons, i.e., $p p \to W \to \ell \nu$. Of course we then need observables
which are sensitive to $m_W$. Two classic examples of such observables are the
transverse mass of the $W$ boson, $m_\perp^{\ell\nu}$, and the transverse
momentum of the charged lepton from the decay of the $W$ boson, $p_\perp^\ell$.
Both observables have the appealing feature that in the absence of higher-order
corrections and with an ideal detector (and very narrow $W$ width) the
distributions of these observables have sharp kinematic edges (at
$m_\perp^{\ell\nu} = m_W$ and $p_\perp^\ell = m_W/2$, respectively) that would
be easy to measure precisely. These edges get washed out by detector effects
in the case of the transverse mass and by higher-order corrections in the case
of the transverse momentum. This makes the two observables complementary since
in the former case the issue is mostly an experimental one while in the latter
case the distribution can be measured very precisely while more involved theory
predictions are necessary to extract $m_W$.

The very high target precision for the $W$ boson mass means that we also have to
reconsider the approach by which we deal with the theory predictions. The
standard tools (collinear factorisation, fixed-order perturbation theory,
resummation, parton showers...) usually let us reach uncertainties at the $1\%$
level. We cannot hope to predict kinematic distributions at a level of
$\mathcal{O}(0.01\%)$ uncertainty from first principles. As a way out, we can
combine measurements of $W$ and $Z$ boson production, parametrise the $Z$
distributions in a QCD-motivated way and transfer them to $W$ distributions,
arguing that the bulk of QCD does not distinguish between $W$ and $Z$ bosons.
This means that we have to focus on modelling effects that do distinguish
between $W$ and $Z$ boson production to the desired level of accuracy. However,
this also implies that we have to take into account effects that were previously
deemed irrelevant. One such example would obviously be electroweak corrections.

\begin{figure}
  \centering
  \begin{subfigure}{0.32\textwidth}
    \begin{tikzpicture}[x=1em,y=1em]
      \node at (0,0)
        {\includegraphics[width=10em]{%
          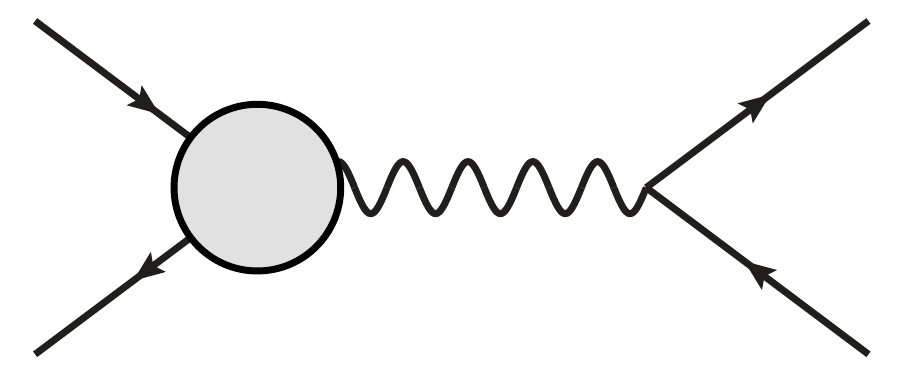}};
      \node[anchor=base] at (-2.1,-0.25) {\small $\alpha$};
      \node at (-5.2, 2.0) {\small $q_a$};
      \node at (-5.2,-2.0) {\small $\bar{q}_b$};
      \node at ( 5.2, 2.0) {\small $\ell_1$};
      \node at ( 5.2,-2.0) {\small $\bar{\ell}_2$};
      \node at ( 0.0,-1.0) {\small $W,Z$};
    \end{tikzpicture}
    \caption{NLO electroweak initial state corrections}
    \label{fig:diag-nlo-ew-is}
  \end{subfigure}
  \begin{subfigure}{0.32\textwidth}
    \begin{tikzpicture}[x=1em,y=1em]
      \node at (0,0)
        {\includegraphics[width=10em]{%
          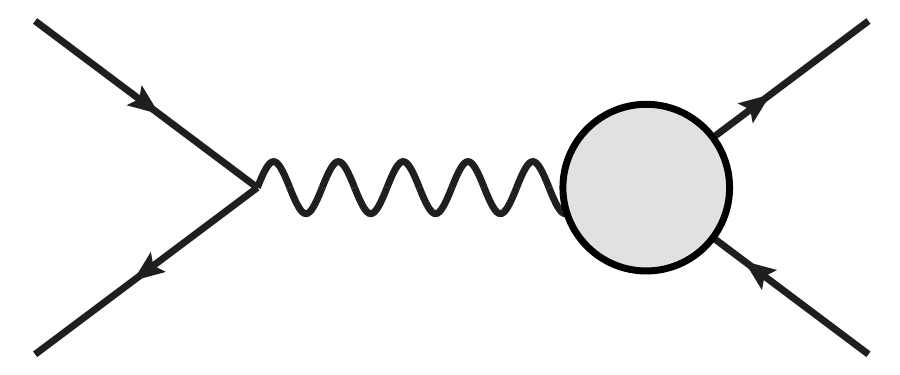}};
      \node[anchor=base] at (2.2,-0.25) {\small $\alpha$};
      \node at (-5.2, 2.0) {\small $q_a$};
      \node at (-5.2,-2.0) {\small $\bar{q}_b$};
      \node at ( 5.2, 2.0) {\small $\ell_1$};
      \node at ( 5.2,-2.0) {\small $\bar{\ell}_2$};
      \node at ( 0.0,-1.0) {\small $W,Z$};
    \end{tikzpicture}
    \caption{NLO electroweak final state corrections}
    \label{fig:diag-nlo-ew-fs}
  \end{subfigure}
  \begin{subfigure}{0.32\textwidth}
    \begin{tikzpicture}[x=1em,y=1em]
      \node at (0,0)
        {\includegraphics[width=10em]{%
          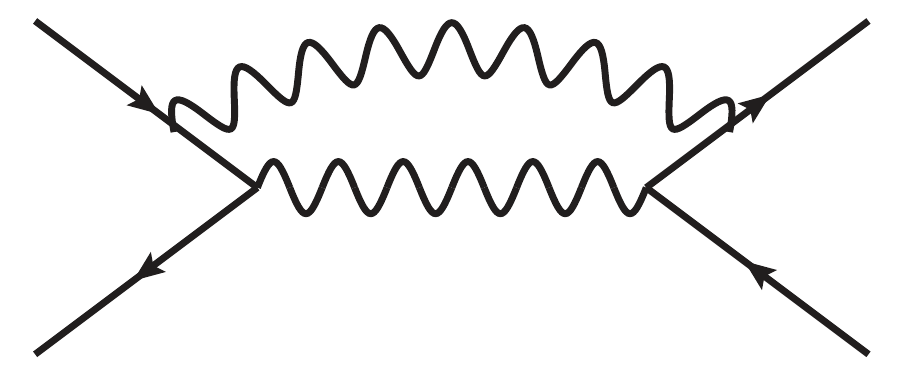}};
      \node at (-5.2, 2.0) {\small $q_a$};
      \node at (-5.2,-2.0) {\small $\bar{q}_b$};
      \node at ( 5.2, 2.0) {\small $\ell_1$};
      \node at ( 5.2,-2.0) {\small $\bar{\ell}_2$};
      \node at ( 0.0,-1.0) {\small $W,Z$};
      \node at ( 0.0, 2.3) {\small $W,Z,\gamma$};
    \end{tikzpicture}
    \caption{NLO electroweak non-factorisable corrections}
    \label{fig:diag-nlo-ew-non-fact}
  \end{subfigure}
  \begin{subfigure}{0.32\textwidth}
    \begin{tikzpicture}[x=1em,y=1em]
      \node at (0,0)
        {\includegraphics[width=10em]{%
          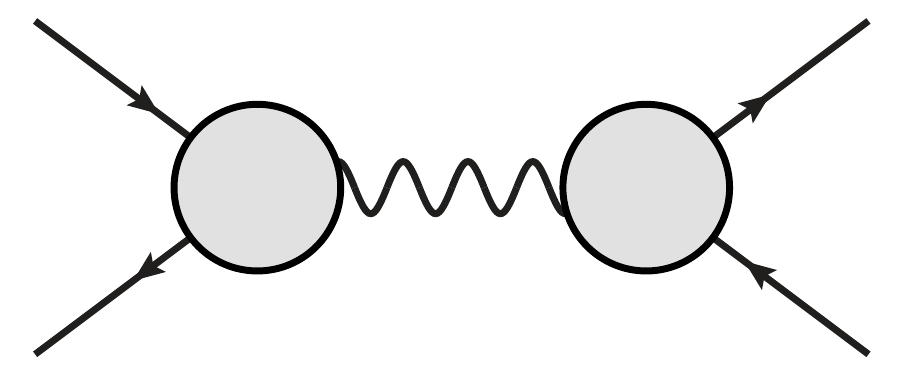}};
      \node[anchor=base] at (-2.1,-0.25) {\small $\alpha_s$};
      \node[anchor=base] at (2.2,-0.25) {\small $\alpha$};
      \node at (-5.2, 2.0) {\small $q_a$};
      \node at (-5.2,-2.0) {\small $\bar{q}_b$};
      \node at ( 5.2, 2.0) {\small $\ell_1$};
      \node at ( 5.2,-2.0) {\small $\bar{\ell}_2$};
      \node at ( 0.0,-1.0) {\small $W,Z$};
    \end{tikzpicture}
    \caption{Mixed QCD-electroweak initial-final corrections}
    \label{fig:diag-mixed-if}
  \end{subfigure}
  \begin{subfigure}{0.32\textwidth}
    \begin{tikzpicture}[x=1em,y=1em]
      \node at (0,0)
        {\includegraphics[width=10em]{%
          figures/blob-fs}};
      \node[anchor=base] at (2.2,-0.25) {\small $\alpha \alpha_s$};
      \node at (-5.2, 2.0) {\small $q_a$};
      \node at (-5.2,-2.0) {\small $\bar{q}_b$};
      \node at ( 5.2, 2.0) {\small $\ell_1$};
      \node at ( 5.2,-2.0) {\small $\bar{\ell}_2$};
      \node at ( 0.0,-1.0) {\small $W,Z$};
    \end{tikzpicture}
    \caption{Mixed QCD-electroweak final-final corrections}
    \label{fig:diag-mixed-ff}
  \end{subfigure}
  \begin{subfigure}{0.32\textwidth}
    \begin{tikzpicture}[x=1em,y=1em]
      \node at (0,0)
        {\includegraphics[width=10em]{%
          figures/blob-is}};
      \node[anchor=base] at (-2.1,-0.25) {\small $\alpha_s\alpha$};
      \node at (-5.2, 2.0) {\small $q_a$};
      \node at (-5.2,-2.0) {\small $\bar{q}_b$};
      \node at ( 5.2, 2.0) {\small $\ell_1$};
      \node at ( 5.2,-2.0) {\small $\bar{\ell}_2$};
      \node at ( 0.0,-1.0) {\small $W,Z$};
    \end{tikzpicture}
    \caption{Mixed QCD-electroweak initial-initial corrections}
    \label{fig:diag-mixed-ii}
  \end{subfigure}
  \caption{Schematic Feynman diagrams illustrating the different types
    of contributions to the single vector boson production process.}
  \label{fig:diags}
\end{figure}
For the goal of measuring $m_W$, we deal with on-shell $W$ and $Z$ boson
production, which allows us to use the narrow width approximation. Since then
the production and decay processes factorise, also the corrections can be
classified as corrections to the initial or the final state. At NLO QCD
corrections can only occur on the initial state, while NLO electroweak
corrections affect both the initial and final state
(\cref{fig:diag-nlo-ew-is,fig:diag-nlo-ew-fs}). In principle, there are also
non-factorising corrections where the initial and final state exchange a $W$,
$Z$ or photon (\cref{fig:diag-nlo-ew-non-fact}), but it has been shown
\cite{Dittmaier:2014qza} that these corrections are suppressed by
$\Gamma_V/m_V$, which parametrically corresponds to another power of $\alpha$.
At the next order of perturbation theory, we can have mixed QCD-electroweak
corrections for the first time. They can again be classified according to which
part of the process receives corrections. The initial-final
(\cref{fig:diag-mixed-if}) and final-final (\cref{fig:diag-mixed-ff})
corrections either factorise into NLO$\otimes$NLO corrections or are only due to
renormalisation contributions, respectively, and they have been dealt with in
Refs.~\cite{Dittmaier:2015rxo,CarloniCalame:2016ouw}. There it was estimated
that they can amount to a shift in the $m_W$ measurement of about
$\SI{15}{\MeV}$. On the other hand, the initial-initial corrections
(\cref{fig:diag-mixed-if}) require a genuine NNLO-type calculation and have not
been known until recently. They have generated a lot of activity, especially in
the in the past few years
\cite{deFlorian:2018wcj,Cieri:2020ikq,Bonciani:2019nuy,Bonciani:2020tvf,%
Dittmaier:2020vra,Heller:2020owb,Buonocore:2021rxx,Bonciani:2021zzf,%
Bonciani:2021iis}, and they are the subject of this proceedings contribution
and the papers on which it is based
\cite{Delto:2019ewv,Buccioni:2020cfi,Behring:2020cqi,Behring:2021adr}.

A number of building blocks are required to complete the initial-initial mixed
QCD-electro\-weak corrections to weak gauge boson production. We need
amplitudes at tree-level with up to two additional emissions of a photon and/or
a gluon, one-loop amplitudes with one emission and finally also the two-loop
form factors for on-shell $W$ and $Z$ bosons. Moreover, since the additional
emissions of massless bosons lead to infrared singularities if those particles
become soft or collinear to other massless partons, we have to use an NNLO
subtraction scheme to isolate the divergences and cancel them against the
divergences from virtual corrections. We adapt the nested soft-collinear
subtraction scheme, which was developed for NNLO QCD calculations, to
QCD-electroweak corrections. In the following, we briefly highlight two aspects
of the calculation, the two-loop on-shell $W$ boson form factor as well as the
subtraction scheme, before discussing the results for the $W$ mass measurements.

\section{Two-loop amplitudes}

Since we work in the narrow-width approximation, we need the two-loop on-shell
form factors for $W$ and $Z$ productions. This is a simplification compared to
the more general off-shell case where also complicated two-loop four-point
functions are required (see
\cite{Bonciani:2016ypc,vonManteuffel:2017myy,Heller:2019gkq,Hasan:2020vwn} for
some recent developments). The mixed QCD-EW corrections to the on-shell form
factor for $Z$ boson production have already been known for over a decade
\cite{Kotikov:2007vr}. For $W$ production, to the best of our knowledge, the
form factor has not been publicly available. Therefore, we have calculated the
missing integrals and completed the form factor.

For the $W$ boson form factor, we have to calculate 44 Feynman diagrams which we
reduce via integration-by-parts relations to 35 master integrals. Of those
integrals 25 are already available in the literature
\cite{Bonciani:2016ypc,Aglietti:2003yc,Aglietti:2004tq}. They all fall into the
category where there is at most one internal mass, i.e., when there are only
either $W$ or $Z$ bosons on internal lines. These integrals are sufficient to
calculate the $Z$ boson form factor. Additionally, the $W$ boson form factor
requires 10 integrals in which $W$ and $Z$ bosons appear on internal lines
simultaneously and which were not available in the literature.

In order to calculate them, we derived differential equations in the mass ratio
$z = m_W^2 / m_Z^2$,
\begin{align}
  \frac{\rmd}{\rmd z} I(z,\epsilon) &= A(z,\epsilon) I(z,\epsilon)
  \,,
\end{align}
and use the equal mass case, $z = 1$, as boundary conditions. The required
boundary constants then fall into the class of integrals that are already
available in the literature.

The differential equation can be solved in terms of GPLs if we use the rational
variable transformation $z = \frac{y}{(1+y)^2}$. An even more compact
representation of the result can be found if we use the original mass ratio
variable $z$ and iterated integrals, defined recursively via
\begin{align}
  H_{a,\vec{b}}(z) &= \int_0^z f_a(t) H_{\vec{b}}(t) \rmd t
  \,, &
  H_{\emptyset}(z) &= 1
  \,, &
  H_{\underbrace{0,\dots,0}_{n\text{ times}}}(z) &= \frac{\ln^n(z)}{n!}
  \,,
\end{align}
over the alphabet
\begin{align}
  f_1(t) &= \frac{1}{1-t}
  \,, &
  f_0(t) &= \frac{1}{t}
  \,, &
  f_{-1}(t) &= \frac{1}{1+t}
  \,, &
  f_{r}(t) &= \frac{1}{\sqrt{t (4-t)}}
  \,.
\end{align}
We presented the result for the form factor in Appendix B of
Ref.~\cite{Behring:2020cqi}.

\section{Subtraction scheme}

Fixed-order calculations beyond tree level develop infrared singularities in
real corrections which lead to $1/\epsilon$ poles after integrating over the
phase space and which cancel against corresponding poles in the virtual
corrections. Since the phase space integration is often carried out
numerically, e.g., to include non-trivial phase space constraints, the infrared
poles have to be extracted and cancelled before numerical integration becomes
possible. A common way to do that is to use subtraction schemes. The basic
idea of this method is to introduce a term that behaves exactly like the
infrared singularity in singular regions of phase space but which can still be
integrated explicitly. By subtracting this term, singularities of real-emission
matrix elements become regulated. This part can be integrated numerically. In
the second part, one adds back the subtracted term and integrates it explicitly
over the singular phase space region, thereby exposing the $1/\epsilon$ poles.
Since we add and subtract the same term, the overall expression stays unchanged.
There has been a lot of progress with NNLO subtraction schemes for QCD over the
past decade. We build on this progress by adapting the nested-soft collinear
subtraction scheme
\cite{Caola:2017dug,Caola:2019nzf,Caola:2019pfz,Asteriadis:2019dte} to mixed
QCD-electroweak corrections. For the $Z$ boson, it is sufficient to take an
implementation of the subtraction scheme for NNLO QCD corrections and to replace
colour factors according to simple abelianisation rules
\cite{deFlorian:2018wcj,Delto:2019ewv}. For the $W$ boson, on the other hand,
new contributions arise due to the fact that $W$ bosons can radiate photons
which in turn gives rise to new types of infrared singularities compared to the
NNLO QCD case.

However, we also profit from simplifications in mixed QCD-electroweak
corrections compared to NNLO QCD. One such example are the triple collinear
limits, which have overlapping singularities in NNLO QCD due to the fact that
the two emitted partons can become collinear at different rates: the parton
emitted earlier can become collinear to the emitting parton faster than the
parton emitted later or vice versa, or they can become collinear to each first
and only then collinear to the emitting parton. In the original formulation of
the nested soft-collinear subtraction scheme, four sectors are introduced to
disentangle these singularities. In the mixed QCD-electroweak case, when a
gluon and a photon are emitted, the collinear limit of these two bosons does not
give rise to a singularity. Therefore, we can drop two sectors and thereby
simplify the construction. Moreover, no new collinear limits arise in the mixed
QCD-electroweak case compared to NNLO QCD.

An even more dramatic simplification occurs in the double-soft limit. At NNLO
in QCD the double-soft eikonal function has non-trivial overlapping
singularities due to the fact that the emitted partons can become soft at
different rates. One way to deal with this is to introduce an energy ordering
by inserting a partition of unity via $1 = \theta(E_{g_1} - E_{g_2}) +
\theta(E_{g_2} - E_{g_1})$, where $E_{g_1}$ and $E_{g_2}$ are the energies of
the two gluons, respectively, and then use different sector decomposition
transformations in each of the two sectors. However, for mixed QCD-electroweak
corrections, soft photons and soft gluons are not entangled and the double-soft
limit factorises into a NLO QCD part and a NLO QED part. As an example, if we
take the squared matrix element for $W$ production with the emission of a gluon
and a photon, $|\mathcal{M}_{Wg\gamma}|^2$, and take the soft limit of the gluon
and photon, we obtain
\begin{align}
  \lim_{E_g,E_\gamma \to 0} |\mathcal{M}_{Wg\gamma}|^2
    \simeq{}& g_s^2 \text{Eik}_g(p_u,p_{\bar{d}};p_g) \,
         e^2 \text{Eik}_\gamma(p_u,p_{\bar{d}},p_W;p_\gamma)
         |\mathcal{M}_W|^2
  \,,
\end{align}
where $|\mathcal{M}_W|^2$ is the squared matrix element for $W$ production and
the QCD soft eikonal function reads
\begin{align}
  \text{Eik}_g(p_u,p_{\bar{d}};p_g)
    &= 2 C_F \frac{(p_u \cdot p_{\bar{d}})}{%
      (p_u \cdot p_g)
      (p_g \cdot p_{\bar{d}})}
  \,.
\end{align}
Thus, there is no need to distinguish whether the gluon or the photon becomes
soft faster and we do not have to introduce an energy ordering. This simplifies
the soft limits tremendously.

The fact that the $W$ boson is electrically charged and, therefore, can radiate
photons, leads to new contributions that do not have a correspondence in the
NNLO QCD calculation. The mass of the $W$ boson screens against collinear
singularities from this photon, but the photon can still become soft and cause a
singularity that way. To construct a subtraction term for this limit, we need
the soft eikonal function for massive emitters. But since QCD and QED factorise
in the soft limit, as discussed above, only NLO eikonal functions are necessary.
The corresponding NLO QED soft eikonal function reads
\begin{align}
  \text{Eik}_\gamma(p_u,p_{\bar{d}},p_W; p_\gamma)
    ={}& \left\{
           Q_u Q_d \frac{2 (p_u \cdot p_{\bar{d}})}{%
             (p_u \cdot p_\gamma) (p_{\bar{d}} \cdot p_\gamma)}
           -Q_W^2 \frac{p_W^2}{(p_W \cdot p_\gamma)^2}
  \right. \notag \\ & \left.
           +Q_W \left(
             Q_u \frac{2 (p_W \cdot p_u)}{%
               (p_W \cdot p_\gamma) (p_u \cdot p_\gamma)}
             -Q_d \frac{2 (p_W \cdot p_{\bar{d}})}{%
               (p_W \cdot p_\gamma) (p_{\bar{d}} \cdot p_\gamma)}
           \right)
         \right\}
  \,.
\end{align}

More details about the subtraction scheme for mixed QCD-electroweak corrections
to on-shell $W$ production are presented in Ref.~\cite{Behring:2020cqi}.

\section{Estimates of the impact of mixed corrections on the
  \texorpdfstring{$W$}{W} mass}

Once all required building blocks are available it becomes possible to calculate
fiducial cross sections and differential distributions for on-shell $W$ and $Z$
boson production. In Refs.~\cite{Buccioni:2020cfi,Behring:2020cqi} it was shown
that mixed QCD-electroweak corrections to these processes are very small, about
$\mathcal{O}(0.05\%)$, but not obviously irrelevant for $m_W$ measurements at
the LHC. Therefore, it is interesting to \emph{estimate} the impact of the
\emph{newly-computed} initial-initial mixed QCD-electroweak corrections on the
$W$ boson mass measurements. There were a number of considerations that guided
us when we devised how to derive these estimates:
\begin{itemize}
  \item The method should combine $W$ and $Z$ measurements, since, as discussed
        above, this models what is being done in experimental analyses and also
        makes use of the precision that is available for the $Z$ mass
        measurements at LEP.
  \item The method should be physically and conceptually simple and transparent.
        The experimental collaborations use an intricate template-fit-based
        method which would require a careful implementation of all relevant
        effects besides the new corrections.
  \item The method should be accessible with our calculations. Since we work in
        fixed-order perturbation theory and employ the narrow-width
        approximation, we cannot use the transverse mass as it is sensitive to
        off-shell effects which are not appropriately captured by this
        description. Instead, we use the transverse momentum distribution of
        the charged lepton below.
\end{itemize}
Let us stress again that the goal here is to derive estimates for the shifts
caused by these particular corrections and not to assess all possible effects or
to propose this method for performing the measurement.

We use the average transverse momentum of the charged lepton from the decay of
the weak gauge boson ($V = W, Z$), calculated according to
\begin{align}
  \avgptl{V}
    &= \frac{\int \rmd \sigma_V \times \ptl}{\int \rmd \sigma_V}
  \,,
\end{align}
where the phase space integration may be subject to constraints, e.g., from
fiducial cuts. Thus, we effectively calculate the first moment of the $\ptl$
distribution normalised to the fiducial cross section. Since the $p_\perp^\ell$
distribution has a peak, which results from the smeared edge at $\ptl =
\frac{m_V}{2}$, the first moment is of course highly sensitive to the value of
$m_V$. At leading order, when the only cut on the final state is a minimal
$\ptl$, the observable takes the form
\begin{align}
  \avgptl{V} &= m_V f\left(\frac{p_\perp^\text{cut}}{m_V}\right)
  \,, &
  \text{with }
  f(r) &= \frac{3}{32} \frac{r (5-8 r^2)}{1-r^2}
          +\frac{15}{64} \frac{\arcsin(\sqrt{1 - 4 r^2})}{(1-r^2) \sqrt{1 - 4 r^2}}
  \label{eq:avgptlLO}
  \,.
\end{align}
This illustrates that there is a strong dependence of the observable on the
vector boson mass. If we ignore the influence of the cut
($f(p_\perp^\text{cut}/m_V)$) the dependence is linear. With this in mind, we
now construct the observable
\begin{align}
  m_W^\text{meas}
    &= \frac{\avgptl{W}^\text{meas}}{\avgptl{Z}^\text{meas}} m_Z C_\text{th}
  \label{eq:mWmeas}
  \,.
\end{align}
The average lepton transverse momenta $\avgptl{V}$ are taken as measurements
from the LHC. As we just discussed, they each are proportional to $m_V$. The
$Z$ boson mass $m_Z$ is taken from the measurements at LEP, making use of the
precision that is available from there. Finally, there is a theoretical
correction factor $C_\text{th}$, which models all the details that distinguish
between $W$ and $Z$ bosons, including, for example, different fiducial cuts in
both cases. The theoretical correction factor can be calculated by solving
\cref{eq:mWmeas} for $C_\text{th}$ and replacing $\avgptl{V}^\text{meas}$ by
theory calculations, i.e.,
\begin{align}
  C_\text{th} &= \frac{m_W}{m_Z} \frac{\avgptl{Z}^\text{th}}{\avgptl{W}^\text{th}}
  \,.
\end{align}
Therefore, if we add a new contribution, like the mixed QCD-electroweak
corrections, to the theory, the value of $C_\text{th}$ changes and hence also
the extracted $W$ boson mass $m_W^\text{meas}$. In order to quantify the size
of the shift induced by the new correction, we have to use
\begin{align}
  \frac{\delta m_W^\text{meas}}{m_W^\text{meas}}
    &= \frac{\delta C_\text{th}}{C_\text{th}}
     = \frac{\delta \avgptl{Z}}{\avgptl{Z}}
       -\frac{\delta \avgptl{W}}{\avgptl{W}}
  \label{eq:deltamW}
  \,.
\end{align}
The last step also highlights the fact that the shift of $m_W$ is sensitive to
differences between the $W$ and $Z$ boson cases: if the observable changes in
exactly the same way in both cases, the change cancels out in the final shift.

\begin{table}
  \caption{Estimates for the shifts of $m_W$ due to the NLO electroweak and
           the mixed QCD-electroweak corrections for different values of the
           factorisation and renormalisation scales $\mu = \mu_R = \mu_F$. For
           details on the three sets of fiducial cuts (``inclusive'',
           ``fiducial'' and ``tunded fiducial'') see the main text.}
  \label{tab:shifts}
  \centering
  \begin{tabular}{llccc}
    \toprule
    $\delta m_W\,[\si{\MeV}]$ & & $\mu = m_V/4$ & $\mu = m_V/2$ & $\mu = m_V$ \\
    \midrule
    \multirow{2}{*}{Inclusive}
      & NLO EW & $-0.1$ & $0.3$ & $0.2$ \\
      & QCD-EW & $-5.1$ & $-7.5$ & $-9.3$ \\
    \midrule
    \multirow{2}{*}{Fiducial}
      & NLO EW & $0.2$ & $2.3$ & $4.2$ \\
      & QCD-EW & $-16$ & $-17$ & $-19$ \\
    \midrule
    \multirow{2}{*}{Tuned fiducial}
      & NLO EW & $-4.4$ & $-2.5$ & $-0.8$ \\
      & QCD-EW & $3.9$ & $-1.0$ & $-5.7$ \\
    \bottomrule
  \end{tabular}
\end{table}
If we use \cref{eq:deltamW} to derive estimates for the shift of the measured
value of the $W$ boson mass due to the inclusion of NLO electroweak or mixed
QCD-electroweak corrections, we find the results presented in table
\cref{tab:shifts}. If we consider the inclusive case, i.e., we do not impose
any cuts, we find shifts of about \SI{-7}{\MeV} for the mixed QCD-electroweak
corrections. These are much larger than the shifts induced by the NLO
electroweak corrections, which are below \SI{1}{\MeV}. One reason for this is
that the NLO electroweak corrections are particularly small due to our choice to
work in the $G_\mu$ input parameter scheme, which is known to reduce the size of
NLO electroweak corrections in Drell-Yan-type processes. The other reason is a
strong cancellation between the changes in the $W$ and the $Z$ boson cases. To
illustrate this point, we have also calculated the shifts that would arise if we
artificially set $\delta \avgptl{Z}$ to zero. Then we would find $\delta m_W
\approx \SI{-31}{\MeV}$ for NLO electroweak corrections and $\delta m_W \approx
\SI{54}{\MeV}$ for mixed QCD-electroweak corrections. Comparing these values to
the results in the table, we see that there is a particularly strong
cancellation between $\avgptl{Z}$ and $\avgptl{W}$ for NLO electroweak
corrections.

The qualitative picture remains the same if we apply fiducial cuts to both the
$W$ and the $Z$ boson observables. However, the size of corrections increases
by about a factor $2.5$. The cuts are inspired by the ATLAS analysis
\cite{ATLAS:2017rzl} and involve, among other things, a cut on the transverse
momentum of the charged leptons: $p_\perp^{e^+} > \SI{30}{\GeV}$ for $W$
production and $p_\perp^{e^\pm} > \SI{25}{\GeV}$ for $Z$ production. This
difference is sufficient to explain the larger size of the shifts in this setup.
The average transverse momentum is sensitive to the ratio $p_\perp^\text{cut} /
m_V$, as can also be seen in \cref{eq:avgptlLO}. However, ATLAS applies larger
cuts on $p_\perp$ for $W$ production than for $Z$ production, while the $Z$
boson is of course heavier than the $W$ boson. This leads to a decorrelation
between the $W$ and $Z$ boson observables and therefore disturbs the
cancellation in \cref{eq:deltamW}.

Since the size of the shifts strongly depends on the fiducial cuts, we can also
try to use this to ``tune'' the cuts such that the size of the mixed
QCD-electroweak corrections is minimised. The results in \cref{tab:shifts} for
the ``tuned fiducial'' setup show that this is indeed possible. We start from
the same cuts as in the ``fiducial'' setup and adjust the cut on $p_\perp^{e^+}$
for $W$ production such that $C_\text{th} = 1$ at leading order. To achieve
this, we have to require $p_\perp^{e^+} > \SI{25.44}{\GeV}$ for $W$ production.
With these cuts, the impact of the mixed QCD-electroweak corrections gets
substantially reduced. This serves to show that fiducial cuts are an important
factor when assessing the impact of mixed QCD-electroweak corrections.

\section{Conclusion}
We have calculated the mixed QCD-electroweak corrections to fully-differential
on-shell $W$ and $Z$ production at the LHC. This became possible due to
progress on amplitude calculations and subtraction schemes. We found that these
corrections are small, in line with the expectations, but they are not obviously
irrelevant for the $W$ boson mass measurements. Since experimental measurements
of $m_W$ rely on the similarity of $W$ and $Z$ distributions, we have built a
simple model to estimate shifts on $m_W$ using the average transverse momentum
of the charged leptons. We find that the mixed QCD-electroweak corrections
induce shifts on $m_W$ that are comparable to or even larger than the target
precision of $\mathcal{O}(\SI{10}{\MeV})$. Moreover, the size of the shifts is
strongly dependent on the fiducial cuts. Further investigations on the impact
of mixed QCD-electroweak corrections to $m_W$ are clearly warranted, and they
should reflect all relevant details of the experimental analyses.

\section*{Acknowledgements}
The author would like to thank F.~Buccioni, F.~Caola, M.~Delto, M.~Jaquier,
K.~Melnikov and R.~Röntsch for the fruitful collaboration on the topic on
which the results in this article are based as well as K.~Melnikov for comments
on the manuscript.

\paragraph{Funding information}
This research is partially supported by the Deutsche Forschungsgemeinschaft
(DFG, German Research Foundation) under grant 396021762 - TRR 257.

\bibliography{proceedings}

\begin{thebibliography}{10}
\providecommand{\url}[1]{\texttt{#1}}
\providecommand{\urlprefix}{URL }
\expandafter\ifx\csname urlstyle\endcsname\relax
  \providecommand{\doi}[1]{doi:\discretionary{}{}{}#1}\else
  \providecommand{\doi}{doi:\discretionary{}{}{}\begingroup
  \urlstyle{rm}\Url}\fi
\providecommand{\eprint}[2][]{\url{#2}}

\bibitem{ATLAS:2017rzl}
M.~Aaboud \emph{et~al.},
\newblock \emph{{Measurement of the $W$-boson mass in pp collisions at
  $\sqrt{s}=7$ TeV with the ATLAS detector}},
\newblock Eur. Phys. J. C \textbf{78}(2), 110 (2018),
\newblock \doi{10.1140/epjc/s10052-017-5475-4},
\newblock [Erratum: Eur.Phys.J.C 78, 898 (2018)],
\newblock \eprint{1701.07240}.

\bibitem{Baak:2014ora}
M.~Baak, J.~C\'uth, J.~Haller, A.~Hoecker, R.~Kogler, K.~M\"onig, M.~Schott and
  J.~Stelzer,
\newblock \emph{{The global electroweak fit at NNLO and prospects for the LHC
  and ILC}},
\newblock Eur. Phys. J. C \textbf{74}, 3046 (2014),
\newblock \doi{10.1140/epjc/s10052-014-3046-5},
\newblock \eprint{1407.3792}.

\bibitem{deBlas:2016ojx}
J.~de~Blas, M.~Ciuchini, E.~Franco, S.~Mishima, M.~Pierini, L.~Reina and
  L.~Silvestrini,
\newblock \emph{{Electroweak precision observables and Higgs-boson signal
  strengths in the Standard Model and beyond: present and future}},
\newblock JHEP \textbf{12}, 135 (2016),
\newblock \doi{10.1007/JHEP12(2016)135},
\newblock \eprint{1608.01509}.

\bibitem{Dittmaier:2014qza}
S.~Dittmaier, A.~Huss and C.~Schwinn,
\newblock \emph{{Mixed QCD-electroweak $\mathcal{O}(\alpha_s\alpha)$
  corrections to Drell-Yan processes in the resonance region: pole
  approximation and non-factorizable corrections}},
\newblock Nucl. Phys. B \textbf{885}, 318 (2014),
\newblock \doi{10.1016/j.nuclphysb.2014.05.027},
\newblock \eprint{1403.3216}.

\bibitem{Dittmaier:2015rxo}
S.~Dittmaier, A.~Huss and C.~Schwinn,
\newblock \emph{{Dominant mixed QCD-electroweak O($\alpha_s\alpha$) corrections
  to Drell\textendash{}Yan processes in the resonance region}},
\newblock Nucl. Phys. B \textbf{904}, 216 (2016),
\newblock \doi{10.1016/j.nuclphysb.2016.01.006},
\newblock \eprint{1511.08016}.

\bibitem{CarloniCalame:2016ouw}
C.~M. Carloni~Calame, M.~Chiesa, H.~Martinez, G.~Montagna, O.~Nicrosini,
  F.~Piccinini and A.~Vicini,
\newblock \emph{{Precision Measurement of the W-Boson Mass: Theoretical
  Contributions and Uncertainties}},
\newblock Phys. Rev. D \textbf{96}(9), 093005 (2017),
\newblock \doi{10.1103/PhysRevD.96.093005},
\newblock \eprint{1612.02841}.

\bibitem{deFlorian:2018wcj}
D.~de~Florian, M.~Der and I.~Fabre,
\newblock \emph{{QCD$\oplus$QED NNLO corrections to Drell Yan production}},
\newblock Phys. Rev. D \textbf{98}(9), 094008 (2018),
\newblock \doi{10.1103/PhysRevD.98.094008},
\newblock \eprint{1805.12214}.

\bibitem{Cieri:2020ikq}
L.~Cieri, D.~de~Florian, M.~Der and J.~Mazzitelli,
\newblock \emph{{Mixed QCD\ensuremath{\otimes}QED corrections to exclusive
  Drell Yan production using the q$_{T}$ -subtraction method}},
\newblock JHEP \textbf{09}, 155 (2020),
\newblock \doi{10.1007/JHEP09(2020)155},
\newblock \eprint{2005.01315}.

\bibitem{Bonciani:2019nuy}
R.~Bonciani, F.~Buccioni, N.~Rana, I.~Triscari and A.~Vicini,
\newblock \emph{{NNLO QCD$\times$EW corrections to Z production in the
  $q\bar{q}$ channel}},
\newblock Phys. Rev. D \textbf{101}(3), 031301 (2020),
\newblock \doi{10.1103/PhysRevD.101.031301},
\newblock \eprint{1911.06200}.

\bibitem{Bonciani:2020tvf}
R.~Bonciani, F.~Buccioni, N.~Rana and A.~Vicini,
\newblock \emph{{Next-to-Next-to-Leading Order Mixed QCD-Electroweak
  Corrections to on-Shell Z Production}},
\newblock Phys. Rev. Lett. \textbf{125}(23), 232004 (2020),
\newblock \doi{10.1103/PhysRevLett.125.232004},
\newblock \eprint{2007.06518}.

\bibitem{Dittmaier:2020vra}
S.~Dittmaier, T.~Schmidt and J.~Schwarz,
\newblock \emph{{Mixed NNLO QCD\texttimes{}electroweak corrections of
  $\mathcal{O}(N_f \alpha_s \alpha)$ to single-W/Z production at the LHC}},
\newblock JHEP \textbf{12}, 201 (2020),
\newblock \doi{10.1007/JHEP12(2020)201},
\newblock \eprint{2009.02229}.

\bibitem{Heller:2020owb}
M.~Heller, A.~von Manteuffel, R.~M. Schabinger and H.~Spiesberger,
\newblock \emph{{Mixed EW-QCD two-loop amplitudes for $q\bar{q} \to
  \ell^+\ell^-$ and $\gamma_5$ scheme independence of multi-loop corrections}},
\newblock JHEP \textbf{05}, 213 (2021),
\newblock \doi{10.1007/JHEP05(2021)213},
\newblock \eprint{2012.05918}.

\bibitem{Buonocore:2021rxx}
L.~Buonocore, M.~Grazzini, S.~Kallweit, C.~Savoini and F.~Tramontano,
\newblock \emph{{Mixed QCD-EW corrections to
  $\boldsymbol{pp\!\to\!\ell\nu_\ell\!+\!X}$ at the LHC}},
\newblock Phys. Rev. D \textbf{103}, 114012 (2021),
\newblock \doi{10.1103/PhysRevD.103.114012},
\newblock \eprint{2102.12539}.

\bibitem{Bonciani:2021zzf}
R.~Bonciani, L.~Buonocore, M.~Grazzini, S.~Kallweit, N.~Rana, F.~Tramontano and
  A.~Vicini,
\newblock \emph{{Mixed strong$-$electroweak corrections to the Drell$-$Yan
  process}}  (2021),
\newblock \eprint{2106.11953}.

\bibitem{Bonciani:2021iis}
R.~Bonciani, F.~Buccioni, N.~Rana and A.~Vicini,
\newblock \emph{{On-shell Z boson production through ${\cal O}(\alpha
  \alpha_s)$}}  (2021),
\newblock \eprint{2111.12694}.

\bibitem{Delto:2019ewv}
M.~Delto, M.~Jaquier, K.~Melnikov and R.~R\"ontsch,
\newblock \emph{{Mixed QCD$\otimes$QED corrections to on-shell $Z$ boson
  production at the LHC}},
\newblock JHEP \textbf{01}, 043 (2020),
\newblock \doi{10.1007/JHEP01(2020)043},
\newblock \eprint{1909.08428}.

\bibitem{Buccioni:2020cfi}
F.~Buccioni, F.~Caola, M.~Delto, M.~Jaquier, K.~Melnikov and R.~R\"ontsch,
\newblock \emph{{Mixed QCD-electroweak corrections to on-shell Z production at
  the LHC}},
\newblock Phys. Lett. B \textbf{811}, 135969 (2020),
\newblock \doi{10.1016/j.physletb.2020.135969},
\newblock \eprint{2005.10221}.

\bibitem{Behring:2020cqi}
A.~Behring, F.~Buccioni, F.~Caola, M.~Delto, M.~Jaquier, K.~Melnikov and
  R.~R\"ontsch,
\newblock \emph{{Mixed QCD-electroweak corrections to $W$-boson production in
  hadron collisions}},
\newblock Phys. Rev. D \textbf{103}(1), 013008 (2021),
\newblock \doi{10.1103/PhysRevD.103.013008},
\newblock \eprint{2009.10386}.

\bibitem{Behring:2021adr}
A.~Behring, F.~Buccioni, F.~Caola, M.~Delto, M.~Jaquier, K.~Melnikov and
  R.~R\"ontsch,
\newblock \emph{{Estimating the impact of mixed QCD-electroweak corrections on
  the $W$-mass determination at the LHC}},
\newblock Phys. Rev. D \textbf{103}(11), 113002 (2021),
\newblock \doi{10.1103/PhysRevD.103.113002},
\newblock \eprint{2103.02671}.

\bibitem{Bonciani:2016ypc}
R.~Bonciani, S.~Di~Vita, P.~Mastrolia and U.~Schubert,
\newblock \emph{{Two-Loop Master Integrals for the mixed EW-QCD virtual
  corrections to Drell-Yan scattering}},
\newblock JHEP \textbf{09}, 091 (2016),
\newblock \doi{10.1007/JHEP09(2016)091},
\newblock \eprint{1604.08581}.

\bibitem{vonManteuffel:2017myy}
A.~von Manteuffel and R.~M. Schabinger,
\newblock \emph{{Numerical Multi-Loop Calculations via Finite Integrals and
  One-Mass EW-QCD Drell-Yan Master Integrals}},
\newblock JHEP \textbf{04}, 129 (2017),
\newblock \doi{10.1007/JHEP04(2017)129},
\newblock \eprint{1701.06583}.

\bibitem{Heller:2019gkq}
M.~Heller, A.~von Manteuffel and R.~M. Schabinger,
\newblock \emph{{Multiple polylogarithms with algebraic arguments and the
  two-loop EW-QCD Drell-Yan master integrals}},
\newblock Phys. Rev. D \textbf{102}(1), 016025 (2020),
\newblock \doi{10.1103/PhysRevD.102.016025},
\newblock \eprint{1907.00491}.

\bibitem{Hasan:2020vwn}
S.~M. Hasan and U.~Schubert,
\newblock \emph{{Master Integrals for the mixed QCD-QED corrections to the
  Drell-Yan production of a massive lepton pair}},
\newblock JHEP \textbf{11}, 107 (2020),
\newblock \doi{10.1007/JHEP11(2020)107},
\newblock \eprint{2004.14908}.

\bibitem{Kotikov:2007vr}
A.~Kotikov, J.~H. Kuhn and O.~Veretin,
\newblock \emph{{Two-Loop Formfactors in Theories with Mass Gap and Z-Boson
  Production}},
\newblock Nucl. Phys. B \textbf{788}, 47 (2008),
\newblock \doi{10.1016/j.nuclphysb.2007.07.018},
\newblock \eprint{hep-ph/0703013}.

\bibitem{Aglietti:2003yc}
U.~Aglietti and R.~Bonciani,
\newblock \emph{{Master integrals with one massive propagator for the two loop
  electroweak form-factor}},
\newblock Nucl. Phys. B \textbf{668}, 3 (2003),
\newblock \doi{10.1016/j.nuclphysb.2003.07.004},
\newblock \eprint{hep-ph/0304028}.

\bibitem{Aglietti:2004tq}
U.~Aglietti and R.~Bonciani,
\newblock \emph{{Master integrals with 2 and 3 massive propagators for the 2
  loop electroweak form-factor - planar case}},
\newblock Nucl. Phys. B \textbf{698}, 277 (2004),
\newblock \doi{10.1016/j.nuclphysb.2004.07.018},
\newblock \eprint{hep-ph/0401193}.

\bibitem{Caola:2017dug}
F.~Caola, K.~Melnikov and R.~R\"ontsch,
\newblock \emph{{Nested soft-collinear subtractions in NNLO QCD computations}},
\newblock Eur. Phys. J. C \textbf{77}(4), 248 (2017),
\newblock \doi{10.1140/epjc/s10052-017-4774-0},
\newblock \eprint{1702.01352}.

\bibitem{Caola:2019nzf}
F.~Caola, K.~Melnikov and R.~R\"ontsch,
\newblock \emph{{Analytic results for color-singlet production at NNLO QCD with
  the nested soft-collinear subtraction scheme}},
\newblock Eur. Phys. J. C \textbf{79}(5), 386 (2019),
\newblock \doi{10.1140/epjc/s10052-019-6880-7},
\newblock \eprint{1902.02081}.

\bibitem{Caola:2019pfz}
F.~Caola, K.~Melnikov and R.~R\"ontsch,
\newblock \emph{{Analytic results for decays of color singlets to $gg$ and $q
  \bar q$ final states at NNLO QCD with the nested soft-collinear subtraction
  scheme}},
\newblock Eur. Phys. J. C \textbf{79}(12), 1013 (2019),
\newblock \doi{10.1140/epjc/s10052-019-7505-x},
\newblock \eprint{1907.05398}.

\bibitem{Asteriadis:2019dte}
K.~Asteriadis, F.~Caola, K.~Melnikov and R.~R\"ontsch,
\newblock \emph{{Analytic results for deep-inelastic scattering at NNLO QCD
  with the nested soft-collinear subtraction scheme}},
\newblock Eur. Phys. J. C \textbf{80}(1), 8 (2020),
\newblock \doi{10.1140/epjc/s10052-019-7567-9},
\newblock \eprint{1910.13761}.

\end{thebibliography}

\nolinenumbers

\end{document}